# Modeling s-t Path Availability to Support Disaster Vulnerability Assessment of Network Infrastructure

Timothy C. Matisziw[1]
Alan T. Murray[1,2]

[1] Center for Urban and Regional Analysis
 The Ohio State University
 1036 Derby Hall, 154 N. Oval Mall
 Columbus, Ohio 43210

[2] Department of Geography
 The Ohio State University

Email: matisziw.1@osu.edu; murray.308@osu.edu

**Abstract:**

The maintenance of system flow is critical for effective network operation. Any type of disruption to network facilities (arcs/nodes) potentially risks loss of service, leaving users without access to important resources. It is therefore an important goal of planners to assess infrastructures for vulnerabilities, identifying those vital nodes/arcs whose debilitation would compromise the most source-sink (s-t) interaction or system flow. Due to the budgetary limitations of disaster management agencies, protection/fortification and planning for the recovery of these vital infrastructure facilities is a logical and efficient proactive approach to reducing worst-case risk of service disruption. Given damage to a network, evaluating the potential for flow between s-t pairs requires assessing the availability of an operational s-t path. Recent models proposed for identifying infrastructure vital to system flow have relied on enumeration of all s-t paths to support this task. This paper proposes an alternative model constraint structure that does not require complete enumeration of s-t paths, providing computational benefits over existing models. To illustrate the model, an application to a practical infrastructure planning problem is presented.

**Introduction**

Assessing networks for infrastructure vulnerabilities is an important component of strategically planning for disruptive events. Vital infrastructure are those facilities (arcs/nodes) that are most crucial in facilitating/ensuring system operation and therefore, their disruption can pose considerable risk to network service. In many network infrastructures, the loss of only one or a few critical facilities can have severe repercussions on the operability of a system, resulting in wide-ranging service disruptions [1-2]. Hence, evaluating the operability of network infrastructures (e.g, military supply lines, highways, electrical power grids, pipelines, telecommunication systems etc.) given intentional or unplanned disruption is a planning priority [3]. One option for enhancing the state of preparedness for a networked system is though hardening or fortifying network facilities to increase their resistance to damage [4-9]. Fortification of vital facilities can be viewed as reducing/eliminating the worst-case risk. Thus, enhancing the resilience of vital facilities to disruptive events is an important preemptive consideration in any disaster recovery plan (DRP) [10]. This is a particularly significant issue when trying to establish DRPs in response to targeted threats, such as terrorist attack. Since terrorist activities are often geared toward inflicting the greatest damage to infrastructure operation with the least effort, taking away the terrorists' best (highest value) options can in effect reduce the burden associated with planning for and recovering from a disaster. Identification of vital facilities is also a critical step in responding to a disaster. Planning for a worst-case disaster can help ensure that adequate resources are in place to address system recovery. Regardless of the situation of concern, budgetary resources available for disaster management planning are often limited and there is a need to prioritize DRP efforts [10]. In other words, it is crucial that the most important facilities for ensuring network functionality be given priority in plan implementation. Network-based infrastructures, however, are complex topologically as well as functionally, and identification of vital facilities is a challenging endeavor. Further contributing to the complexities of this important planning problem is that many criteria exist for assessing the importance of infrastructure components, as do many approaches for modeling system performance.

In the following section, methodologies for examining facility significance within a system are discussed. Among these methodologies, those that address the preservation of existing network activity, or system flow, are highlighted given their importance when planning for potential loss of service. Next, existing approaches for assessing the potential for loss of system flow are reviewed. In order to better deal with computational issues surrounding these approaches, a new model is proposed to ascertain the survivability of system flow given network disruption. An evaluation of Ohio's interstate highway network is then detailed to illustrate the benefits of the proposed formulation. Finally, discussion and conclusions are provided.

**Background**

The importance of network infrastructure to associated activity is dependant on the location and role of nodes and arcs within a system. Therefore, how topological relationships between network facilities are considered in vulnerability analyses is a critical issue. One way of assessing facility importance within a network is to model the worst-case impact to network performance provided some assumptions on the type of disruptive event expected. Facilities associated with a worst-case scenario are suggestive of their importance to network operation. However, central to this type of analysis is that the disruptive potential of an event can be measured in a variety of ways and there are many levels of disruption that can occur. First, disruptive events in networks typically involve nodes, arcs, or any combination thereof. Thus, a critical issue is how network performance is impacted by damage to arcs and nodes. There are many operational characteristics of networks that can be considered important to system performance. For instance, level of system connectivity/flow, maximum flow capacity, and cost of transportation or service can all be affected by facility damage. The importance of an arc/node to any of these performance measures depends on the availability of other facilities and the topology of the system. Further complicating matters is that the importance of network facilities can also be measured by how their loss indirectly impacts other system interactions. For instance, loss of service between two nodes can additionally impact the ability of the two nodes to fulfill interactions with other connected nodes. Such a case occurs when assessing how regional economic activities are impacted by disruptions to transportation infrastructure [5,7,11]. Another critical issue relates to the level of disruption expected from an event. The disruption level may be represented by the number of facilities anticipated to be involved or the cost associated with removing operational capacity or decreasing operational efficiency. Facilities that are vital to network operation at one assumed level of disruption (e.g., a two node disruption) may not necessarily coincide with those vital at another level (e.g., a four node disruption). Given the myriad of considerations available for assessing network vulnerability, interpretations of vulnerability can be subject to a wide range of variability.

Network optimization methods have received considerable attention in the search for facilities vital to network operation. A variety of approaches have been developed to assess network vulnerabilities given some interpretation of facility importance to system performance. Many optimization models, also known as interdiction models, have been developed to identify important facilities with regard to impact on system service cost [6,12-17], flow capacity [18-23], as well as system connectivity and flow [24-26]. These methods can be used to identify bounds for network vulnerability in terms of facilities associated with worst-case (or best-case) impacts to system performance. Constraints on the amount of resources available for disruption of network cost, capacity, or system flow are often stipulated in accordance with assumptions related to the type of damage anticipated. Using such techniques for establishing vulnerability bounds provides extremely useful insights for strategic planning efforts, such as DRP development, geared at protecting or fortifying critical network facilities.

Of particular interest in this paper is identifying network facilities most vital to system flow. It is assumed in this context that there are a number of sources and sinks (s-t) of flow within a network and that they interact at a specified level. This interaction or flow between s-t pairs $(f_{st})$ is the measure of system performance considered vulnerable to disruption. From a disaster management perspective, this is the activity or service at risk and possibly needing restoration post-disaster. Here, the level of interaction between two nodes is considered independent of the level of interaction between other nodal pairs.

It is also assumed that s-t activity is supported by the availability of operational s-t paths and that connectivity can be achieved given the presence of an operational path, regardless of its characteristics. As discussed earlier, identifying the worst-case disaster outcomes can contribute significant insight to the disaster planning process. Several optimization methods have been proposed to identify the nodes or arcs associated with the worst-case impact to system flow, given restrictions on the number of facilities damaged. In this context, facility damage equates to the complete debilitation of individual components. Myung and Kim [24] propose an approach to assess the survivability of a network to facility disruption. To address this concern, they develop an integer program to maximize, or provide an upper bound on, the total flow rendered incapable of interaction, given the loss of *p* arcs. For an uncapacitated network *G* with node set *N* and arc set *A* (*G=(N,A)*), Myung and Kim [24] formulate their survivability model as follows:

Notation:

$s \in N$ = index of sources

$t \in N$ = index of sinks

$i \in A$ = index of network arcs

$f_{st}$ = flow observed between *s-t*

$\beta \in B$ = index of paths

$N_{st}$ = set of paths capable of establishing s-t connectivity

$p$ = number of arcs to be disabled

$X_i = \begin{cases} 1, \text{if arc } i \text{ is removed} \\ 0, \text{otherwise} \end{cases}$

$Z_{st} = \begin{cases} 1, \text{if no path exists between } s,t \\ 0, \text{otherwise} \end{cases}$

$$Maximize \sum_s \sum_t f_{st} Z_{st} \qquad (1)$$

s.t.

$$\sum_i X_i \leq p \qquad (2)$$

$$\sum_{i \in A(\beta)} X_i \geq Z_{st} \qquad \forall s, t, \beta \in N_{st} \qquad (3)$$

$$X_i = \{0,1\} \qquad \forall i \qquad (4)$$
$$Z_{st} = \{0,1\} \qquad \forall s, t$$

Objective (1) maximizes system flow disrupted. Constraint (2) restricts the number of facilities lost to $p$. Constraints (3) state that every path connecting a s-t pair must be impacted by facility damage before s-t connectivity can be considered lost. Finally, Constraints (4) reflect the binary nature of the decisions to be made. Heuristic procedures are suggested by Myung and Kim [24] for establishing upper and lower bounds associated with this model formulation. Murray et al. [25] further consider the possibility that both an exact upper and lower bound on flow loss may be of interest. An integer program, the Flow Interdiction Model (FIM), is developed to establish these bounds and identify associated vital facilities.

The models proposed by both Myung and Kim [24] and Murray et al. [25] rely upon specification of s-t paths as model input. In other words, all potential paths of movement between s-t pairs (and the facilities involved in those paths) are explicitly tracked in the proposed models (e.g., Constraints (3)). If a component facility of a path is damaged, then that path is no longer available to that s-t pair. If there are no s-t paths available due to facility damage, then interaction or flow between the pair is disrupted. For example, Figure 1 shows a network with a single s-t pair. In this network, there are 4 paths connecting s and t: 1) s-3-1-t, 2) s-3-2-t, 3) s-2-t, and 4) s-2-3-1-t. If node 1 is destroyed, 2 s-t paths remain operational. If nodes 3 and 1 are destroyed, one s-t path is still available. However, if nodes 1 and 2 are disabled, then no s-t paths exist and s-t flow is obviously disrupted, corresponding to $Z_{st} = 1$. The system flow impacted is the sum of all s-t flows disrupted $\left( \sum_s \sum_t f_{st} Z_{st} \right)$, and is the quantity that is to be maximized. The premise behind system flow approaches is that if a cutset of nodes/arcs between a s-t pair is disrupted, then connectivity or flow between that s-t pair is disrupted. From a budgetary perspective, damage to a minimum cutset inflicts the most damage to s-t interaction with the greatest efficiency. These path-based models are extremely flexible since an analyst has control over which s-t paths are significant to specific planning applications. A significant issue with models that involve tracking feasible paths between network sources and sinks is that as network size increases, the number of possible s-t paths can quickly become unmanageable. For instance, in the 23 node, 34 arc example presented later in this paper, hundreds of s-t paths exists for each s-t pair, giving nearly 60,000 paths and an equal number of Constraints (3).

Identifying approaches for reducing model size is a necessary component of addressing many large network optimization problems. Often, much of a model's size is attributable to ensuring that network structure is adequately represented. In the context of modeling impacts to system flow, techniques have been developed to help reduce the complexities associated with accounting for all network paths. For instance, Myung and Kim [24] present a preprocessing approach for determining beforehand whether an s-t pair could potentially be impacted by the level of damage specified. Only s-t pairs capable of being disrupted are then considered in the model and, as a result, less paths need be to considered. Additionally, path reduction strategies can be applied to dramatically eliminate paths in many instances. For example, Murray et al. [25] show that in many cases redundant or equivalent paths exist and their removal can decrease computational requirements without jeopardizing optimality. Another way of reducing the computational burden associated with accounting for s-t path availability is to remove paths from consideration that are *unlikely* to impact model decisions. Typically, these approaches involve setting a threshold on the length of paths retained for analysis. For instance, the importance of arcs to s-t connectivity is likely in many cases to be sufficiently accounted for in shorter s-t paths, and hence inclusion of longer, circuitous paths may be unnecessary. Applications of such reduction strategies can be found in models of distributed systems [27], supply networks [28], hub networks [29] and other transportation systems [30]. Although faster solutions times, and in many cases, optimal solutions can be obtained from this type of variable reduction strategy, optimality cannot be guaranteed [28,30-32]. Regardless of the model reduction strategy employed, the number of paths remaining may nonetheless present computational difficulties, especially as network size increases. Given model limitations, ways of coping with the intricacies associated with larger network applications need to be explored. The remainder of this paper proposes one such approach.

Figure 1. Single s-t network example

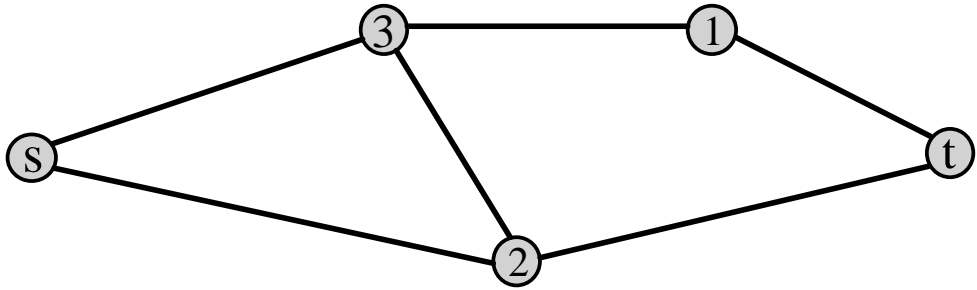

**Path Aggregation in Vulnerability Modeling**

In order to address the computational challenges associated with path-based models, this paper develops an alternative constraint structure for establishing an upper bound on worst-case network disruptions. The model is designed for efficient computation of an upper bound on system connectivity or flow loss and is equivalent to path-based models involving full specification of all s-t paths. Given this equivalency, exact solutions are easily obtained. The proposed constraints act to summarize s-t movement possibilities and do not require the a priori identification of all paths, so fewer constraints are needed. The model is formulated to evaluate disruption to directed arcs, but can easily be modified to consider node disruption. This alternative flow interdiction model incorporating path aggregation constraints (PAC) can be formulated as follows.

Additional notation:

$i, j, k, m \in N$ = indices of nodes

$N_i$ = set of facilities directly connected to $i$

$N_j$ = set of facilities directly connected to $j$

$$X_{ij} = \begin{cases} 1 \text{ if arc } i,j \text{ is disrupted} \\ 0 \text{ otherwise} \end{cases}$$

$$Z_{ij} = \begin{cases} 1 \text{ if a connection between nodes } i \text{ and } j \text{ is unavailable} \\ 0 \text{ otherwise} \end{cases}$$

$$Maximize \sum_s \sum_t f_{st} Z_{st} \qquad (5)$$

s.t.

$$\sum_i \sum_{j \in N_i} X_{ij} = p \qquad (6)$$

$$X_{ij} - Z_{ij} \geq 0 \quad \forall i, j \in N_i \qquad (7)$$

$$X_{ik} + X_{kj} - Z_{ij} \geq 0 \quad \forall i, j, k \in N_i \cap N_j \qquad (8)$$

$$X_{ik} + Z_{km} + X_{mj} - Z_{ij} \geq 0 \quad \forall i, j, k \in N_i, m \in N_j \qquad (9)$$

$$X_{ij} = \{0,1\} \quad \forall i, j \in N_i \qquad (10)$$

$$Z_{ij} = \{0,1\} \quad \forall i, j$$

As in Myung and Kim [24] and Murray et al. [25], objective (5) maximizes total system flow disrupted, and is exactly as specified in (1). Constraint (6) stipulates the number of facilities to be disrupted. Constraints (7)-(9) state that $i$-$j$ flow cannot be disrupted unless no possibility for $i$-$j$ movement exists. More specifically, Constraints (7) consider movement between two adjacent nodes. If a direct link between these nodes is present, then connectivity between the pair ($Z_{ij}$) cannot be compromised. When two nodes are separated by a single, adjacent node (a 2-step connection), Constraints (8) state that if both intervening arcs are available, then connectivity between $i$-$j$ ($Z_{ij}$) has to exist. Essentially, both Constraints (7) & (8) act to account for all paths less than two steps in length. Constraints (9) deal specifically with possible routings where 2 or more intervening nodes occur between an $i$-$j$ pair (more than a 2-step connection). These are the path aggregation constraints (PAC), which account for the availability of all paths greater than two steps in length. In this case, if an arc between $i$ and a directly connected node $k$ ($X_{ik}$), an arc between $j$ and a directly connected node $m$ ($X_{mj}$), and a path between $k$ and $m$ ($Z_{km}$) are available, then there is connectivity between $i$ and $j$ ($Z_{ij}$). Since the model objective maximizes the weighted sum of $Z_{st}$, then $Z_{ij}$ will seek to be one whenever possible. Of course, $Z_{ij}$ can only be disrupted given the absence of all paths (one, two, and multistep) between *i-j* pair. Finally, Constraints (10) impose binary integer restrictions on decision variables.

In this formulation, all paths do not need to be directly specified because all movement possibilities (involving one or many steps) are accounted for collectively in Constraints (7)-(9). This constraint structure works by summarizing *i-j* movement options by defining the necessary components of *i-j* paths, the beginning and ending arcs as well as connectivity between the two. Three main types of paths can occur: 1) 1-step or direct *i-j* connections, 2) 2–step or connections involving an intermediate node adjacent to both the beginning and ending nodes, and 3) multi-step paths involving more than 2 steps. Paths of type 1 and 2 are easily defined explicitly for each *i-j* relationship and require a nominal number of constraints. However, for connections involving more than one intervening node (type 3) numerous paths may be involved. It is in this case where the PAC constraints eliminate the need for path enumeration. The PAC are recursive constraints that act to summarize a set of movement possibilities. That is, each decision on *i-j* connectivity is influenced by the connectivity of other *i-j* pairs. Unless all movement possibilities summarized by the PAC are affected by a disaster, *i-j* movement cannot be disrupted. This is done by using a single PAC constraint representative of all *i-j* paths beginning and ending with the same two arcs. Here this is accomplished by tracking the availability of the beginning and ending arcs as well as the presence of connectivity between the two. Ascertaining whether connectivity exists between the arcs is then derived as a byproduct of evaluating connectivity for other *i-j* pairs. Figure 1 is used to illustrate this concept. In this example, establishing the availability of a connection between s-t involves the existence of at least a two step connection. Specifically, potential for movement between s-t is tracked by:

- Constraints (7): $X_{s\_3} - Z_{s\_3} \geq 0$, $X_{s\_2} - Z_{s\_2} \geq 0$, $X_{2\_3} - Z_{2\_3} \geq 0$, $X_{3\_1} - Z_{3\_1} \geq 0$, $X_{3\_2} - Z_{3\_2} \geq 0$, $X_{2\_t} - Z_{2\_t} \geq 0$, and $X_{1\_t} - Z_{1\_t} \geq 0$

- Constraints (8): $X_{s\_2} + X_{2\_t} - Z_{s\_t} \geq 0$ and $X_{2\_3} + X_{2\_1} - Z_{2\_1} \geq 0$
- Constraints (9): $X_{s\_3} + Z_{3\_1} + X_{1\_t} - Z_{s\_t} \geq 0$,

    $X_{s\_3} + Z_{3\_2} + X_{2\_t} - Z_{s\_t} \geq 0$,

    and $X_{s\_2} + Z_{2\_1} + X_{1\_t} - Z_{s\_t} \geq 0$.

Therefore, if the arcs associated with node 3 are disabled, then constraints $X_{s\_2} - Z_{s\_2} \geq 0$, $X_{2\_t} - Z_{2\_t} \geq 0$, and $X_{s\_2} + X_{2\_t} - Z_{s\_t} \geq 0$ still ensure the availability of an unobstructed s-t connection ($Z_{st}=0$) via node 2. Again, the benefit to structuring the model in this way is that complete paths need not be identified for each s-t pair, and only direct nodal connections are required input. In path-based models, movement possibilities are independently considered for each s-t pair and a set of path constraints is needed for each s-t pair (of which there can be many). The number of these paths can grow exponentially with network size. On the other hand, the PAC rely on establishing the presence of connectivity between all nodes and a smaller set of constraints is needed instead for each *i-j* pair. Given the degree of each node *k*, this set of *i-j* constraints includes at maximum: 1) |N|*|N| of Constraints (7), 2) $\sum_i k_i$ of Constraints (8), and 3) $\sum_i \sum_j k_i * k_j$ of Constraints (9) if a fully connected network is considered. At first glance, for this single s-t example it may appear that there are few computational benefits to this model formulation over its path-based counterparts. If one considers every pair of nodes in this network as source-sink pairings, then 64 s-t paths are required versus 60 constraints using PAC. However, the structural benefits of the PAC become more readily apparent for large networks where there are many s-t pairs and where the number of possible s-t paths can increase exponentially with network size. These computational advantages are highlighted in the next section.

**Ohio's Vital Interstate Infrastructure**

The network vulnerability model of Myung and Kim [24] (eq. (1)-(4)) (referred to M&K from this point on) as well as the equivalent PAC version developed here (eq. (5)-(10)) are applied to address the vulnerability of Ohio's trucking activity to interstate highway disruption. In Figure 2, this portion of the nation's interstate system is shown along with 15 metropolitan statistical areas (MSA), representing generators and attractors of highway traffic within Ohio. In total, this network is composed of 23 vertices (15 of which represent MSAs) and 34 undirected edges. Here inter-MSA linkages are modeled as undirected edges for illustrative purposes (directed arcs could also be considered). For our purposes, trucking between the MSAs is considered to be the measure of network performance vulnerable to network disruption, and hence, damage to network linkages is of concern. Given the focus on inter-MSA edges, the intricacies of the interstate system within each MSA are not considered. Trucking flow data for the year 2000 was obtained from the Ohio Department of Transportation and was aggregated to the MSA level. Given that each of the 15 MSAs interacts with all other MSAs, there are 210 source-sink pairs in this network that can potentially be impacted by edge disruption.

Figure 2. Ohio's interstate system

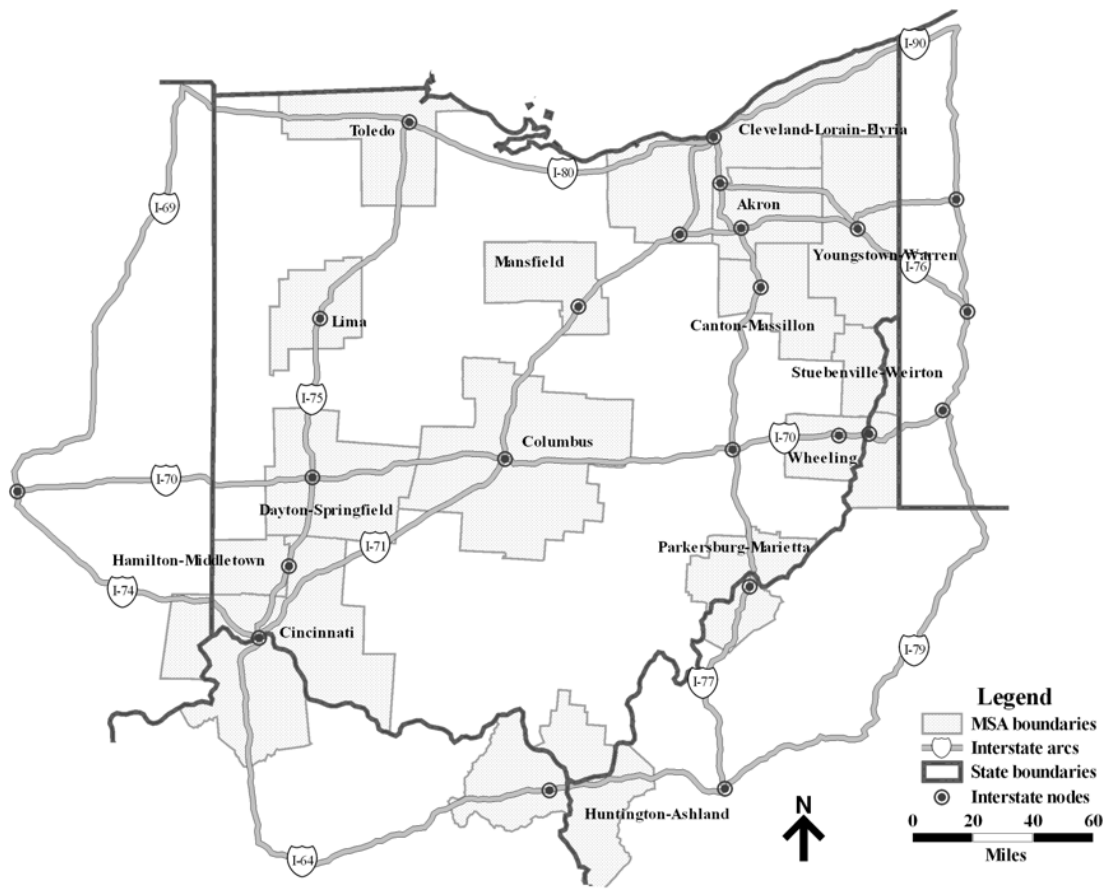

Both the M&K model and PAC version are generated for the network shown in Figure 2 using C++ code and are solved for all levels of edge disruption ($p$=1-34) using ILOG's CPLEX 10.01 mixed integer program solver (on a Xeon 3.0GHz processor with 4GB RAM). For the M&K model, 59,791 simple s-t paths need to be specified a priori. The enumeration and redundancy reduction required to identify these paths consumes nearly 30 minutes of computational effort alone. Additionally, each of these paths then translates into a model constraint. In contrast, only 2,074 constraints are required when employing PAC constraints. This requires approximately 4 constraints per *i-j* pair, whereas in path-based formulations hundreds of paths exist for each s-t pair. For example, assessing connectivity between Toledo and Lima requires only 3 constraints to be specified:

1. $X_{Toledo\_Lima} - Z_{Toledo\_Lima} \geq 0$
2. $X_{Toledo\_Cleveland} + Z_{Cleveland\_Dayton} + X_{Dayton\_Lima} - Z_{Toledo\_Lima} \geq 0$
3. $X_{Toledo\_I69/70 Intersection} + Z_{I69/70 Intersection\_Dayton} + X_{Dayton\_Lima} - Z_{Toledo\_Lima} \geq 0$

Of course, additional constraints are then needed to determine Cleveland-Dayton, I69/70 Interchange-Dayton connectivity, and so on. On the other hand, in the M&K formulation using all paths for Toledo-Lima flow, 321 paths need to be accounted for, some up to 20 steps long.

All 34 problem instances were solved to optimality. All solution results are identical to those obtained using the M&K formulation, identifying the combination of $p$ edges that if damaged, would result in the maximum service disruption. For example, the results indicate that 25 disabled edges would disrupt 100% of network s-t activity. Although damage to a single edge cannot disconnect any MSA from the interstate system ($p$=1), simultaneous disruption of 2 or 3 edges ($p$=2 or 3) can impact over 18% of total daily truck traffic. In either case the solution involves a cutset resulting in the loss of interstate access for the Hamilton MSA. If four edges ($p$=4) are simultaneously lost in a disaster, then over 37% of s-t activity could potentially be impacted. As shown in Figure 3, this involves edge debilitating incidents to I-80 between Toledo and Cleveland, I-70 between Dayton and Columbus, I-75 between Hamilton and Cincinnati, and I-74 between Indianapolis and Cincinnati. Notice these 4 edge form a cutset disconnecting Hamilton, Dayton, Lima, and Toledo from the Ohio's interstate system and the remaining MSAs.

Though the model results produced by the M&K and PAC varieties are equivalent, solution times and computational effort are not. On average only a little over one second was needed to solve each instance using PAC constraints (see Table 1). This is in stark contrast with the average solution time of 84 seconds needed to solve the more heavily constrained path-based version. In terms of computational effort involved in model solution, in the PAC models very little overall effort was required (see Table 1), with an average of 898 iterations and 6 branches involved. In contrast, an average of 28 branches and nearly 4,500 iterations were needed to solve the M&K model.

Figure 3. Vital infrastructure associated with disruption of 4 edges ($p=4$)

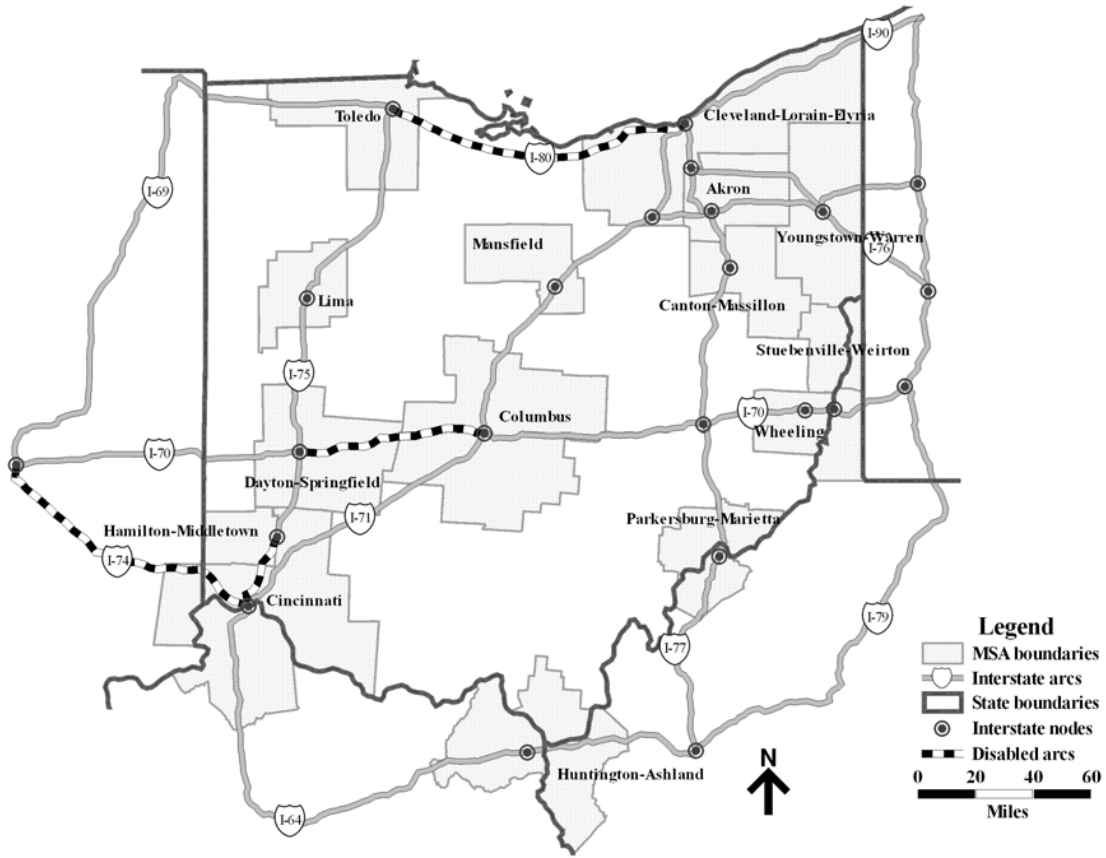

Table 1. Summary model solutions

| p | objective* | PAC** time (sec) | PAC branches | PAC iterations | M&K*** time (sec) | M&K branches | M&K iterations |
|---|---|---|---|---|---|---|---|
| 1 | 0.0 | 4.844 | 26 | 6,073 | 256.641 | 41 | 5,865 |
| 2 | 22,014.5 | 3.172 | 0 | 1,482 | 106.766 | 2 | 3,974 |
| 3 | 22,014.5 | 7.859 | 85 | 7,094 | 243.438 | 60 | 8,801 |
| 4 | 44,151.0 | 5.375 | 4 | 2,602 | 105.984 | 4 | 3,896 |
| 5 | 48,008.0 | 7.36 | 36 | 5,518 | 216.188 | 37 | 6,977 |
| 6 | 75,145.0 | 0.094 | 0 | 299 | 53.297 | 0 | 3,945 |
| 7 | 80,343.5 | 0.265 | 0 | 807 | 92.187 | 2 | 4,120 |
| 8 | 88,109.5 | 0.109 | 0 | 331 | 53.688 | 0 | 4,689 |
| 9 | 93,308.0 | 0.094 | 0 | 316 | 53.438 | 0 | 3,962 |
| 10 | 96,333.0 | 2.11 | 2 | 758 | 73.454 | 4 | 3,608 |
| 11 | 99,234.0 | 2.375 | 18 | 1,230 | 82.532 | 14 | 3,442 |
| 12 | 103,275.5 | 1.437 | 0 | 512 | 77.36 | 12 | 3,317 |
| 13 | 106,300.5 | 0.203 | 0 | 537 | 69.922 | 7 | 3,561 |
| 14 | 109,298.5 | 0.156 | 0 | 432 | 57.703 | 2 | 3,399 |
| 15 | 111,409.0 | 0.14 | 0 | 359 | 59.594 | 2 | 3,618 |
| 16 | 113,124.5 | 0.093 | 0 | 218 | 58.062 | 2 | 4,207 |
| 17 | 114,687.5 | 0.156 | 0 | 282 | 58.25 | 2 | 3,828 |
| 18 | 115,265.5 | 0.765 | 0 | 290 | 74.032 | 36 | 4,074 |
| 19 | 115,973.0 | 0.172 | 0 | 245 | 68.125 | 11 | 4,122 |
| 20 | 116,409.0 | 0.593 | 3 | 272 | 70.5 | 23 | 4,435 |
| 21 | 116,727.5 | 0.094 | 0 | 150 | 88.5 | 106 | 4,933 |
| 22 | 116,868.0 | 0.094 | 0 | 216 | 98.437 | 219 | 5,421 |
| 23 | 116,932.5 | 0.219 | 9 | 251 | 101.157 | 247 | 5,850 |
| 24 | 116,957.5 | 0.203 | 6 | 101 | 77.063 | 106 | 5,684 |
| 25 | 116,958.0 | 0.047 | 0 | 18 | 63.031 | 9 | 4,855 |
| 26 | 116,958.0 | 0.047 | 0 | 26 | 59.593 | 6 | 4,819 |
| 27 | 116,958.0 | 0.047 | 0 | 17 | 54.031 | 0 | 5,413 |
| 28 | 116,958.0 | 0.031 | 0 | 16 | 53.969 | 0 | 4,951 |
| 29 | 116,958.0 | 0.047 | 0 | 17 | 53.953 | 0 | 5,542 |
| 30 | 116,958.0 | 0.047 | 0 | 18 | 53.781 | 0 | 6,878 |
| 31 | 116,958.0 | 0.031 | 0 | 19 | 53.468 | 0 | 4,978 |
| 32 | 116,958.0 | 0.032 | 0 | 17 | 54.125 | 0 | 5,707 |
| 33 | 116,958.0 | 0.031 | 0 | 0 | 97.079 | 0 | 0 |
| 34 | 116,958.0 | 0.016 | 0 | 0 | 0.86 | 0 | 0 |

* Trucks per day impacted

**Results using PAC formulation

***Results using Myung and Kim [24] formulation

**Discussion and Conclusions**

Assessing networks for vulnerabilities is a challenging planning goal spanning many application areas and is of obvious interest in the development of any disaster response plan. Many optimization models have been proposed to facilitate the search for vulnerabilities by identifying nodes and arcs vital to network operation. This variety of methods is essential given that disruption to network operation can be hypothesized to impact network operation in a number of different ways (e.g., loss of capacity, connectivity, efficiency, etc.). For many types of networks, vulnerability relates to how existing patterns of activity or interaction (system flow) is affected by facility damage. Recently, several models have been developed to detect facilities whose loss put the most system flow at risk and do so to generate bounds on network vulnerability [24-25].

Identification of vital network facilities, as detailed in this paper for the Ohio Interstate system, has many potential uses in the disaster planning process. For example, if four transportation links are rendered inoperable, then from a disaster planning perspective protection/fortification of the four edges in Figure 3 could effectively eliminate/decrease the worst-case disruption to inter-MSA truck traffic. Given particular disaster scenarios of interest (and the critical facilities involved), planning agencies could work to devise plans for effective restoration of service for worst-case scenarios. As a range of possible system impacts and recovery strategies might be considered, methodologies for selecting optimal response plans could be employed (see [10]). Aside from assessing protection and recovery options, identification of vital arcs/edges could be used to inform the strategic siting of rapid response and repair stations to facilitate recovery of essential network components. Alternatively, after a disaster, sets of functional network components vital to post-disaster network operation could be identified. For example, one potential goal might be to ensure that remaining infrastructure is not further degraded due to unplanned use. In such a case, vital facilities could be identified in accordance to the level of arc/node management resources dedicated to this goal. A similar use could entail the planning and monitoring of evacuation networks. Planning such a network may involve comparison of several route systems based upon which one is most robust (e.g., minimizes the maximum impact to flow) to various types of disruptions. Monitoring applications could include ensuring that vital nodes/arcs are adequately patrolled to reduce network impedance and facilitate flow.

Thus far, the models proposed for assessing impacts to system flow have focused on tracking potential s-t paths of movement. As discussed earlier, path-based models for assessing network vulnerabilities are advantageous from numerous perspectives. These approaches are ideal for analysis of impacts to system operation given that they can deal with many sources and sinks and do not require networks to be source-sink planar. The primary advantage to these models lies in identifying which s-t paths are vulnerable. Furthermore, path-based models are important given their ability to easily identify a lower bound on network disruption [25]. This aspect is also vital to DRP development since the ability to minimize system flow loss actually provides a means for prioritizing facility repair/restoration/response efforts following a disaster. For instance, given a budget for facility restoration in a particular time frame, minimizing system flow loss can provide the set of facilities that can be restored to support or serve the most s-t flow. However, modeling approaches involving specification of all s-t paths can quickly become computationally prohibitive given the combinatorics involved.

An alternative model structure to the path-based approach, utilizing Path Aggregation Constraints (PAC), was developed here for use in flow vulnerability/survivability models. This approach relies on the fact that only 1 and 2-step *i-j* paths need to be explicitly specified. This approach requires fewer constraints than alternative formulations and improves model solvability, doing so without altering the intent of the model. Given their ability to generalize multi-step paths using beginning and ending arcs as well as intermediate connections, PAC constraints present an extremely useful approach for applications where the number of s-t paths presents computational difficulties. Furthermore, very little effort is needed in identifying PAC constraints versus complete path enumeration. Additionally, the PAC structure is not limited to assessment of arc disruption as presented here as a node-based version is a straightforward extension. Such a modification can be formed by replacing each arc variable with a separate variable for each node involved in that arc. For example, Constraints (7) are then written as: $X_i + X_j - Z_{ij} \geq 0$. An application to truck-based interstate transportation in Ohio illustrated the computational benefits of this model formulation. As shown in this application, the use of the PAC formulation presented a 99% savings in solution time and over an 80% reduction in computational effort (model iterations) involved over alternative approaches. Given its computational benefits, the PAC model structure represents another useful and efficient tool for the disaster planning and restoration efforts.

Though the PAC formulation has computation benefits over existing system flow models, it is not without its limitations. As discussed previously, path-length thresholds have been applied to reduce model size in related system flow approaches. For example, an s-t connection could be considered disconnected if all paths less than a given distance or travel time threshold are impacted by a disaster. Such thresholds cannot be applied in the PAC formulation since the availability of a $Z_{ij}$ connection may involve multiple paths. Therefore, application of the PAC constraints assumes that all network paths are viable for s-t interaction. A related assumption when using system flow models is that any s-t connection can equivalently provide s-t service. However, some instances may arise where connectivity may not only be a function of physical connectivity but also of transportation cost or available capacity. Given that paths are no longer explicitly tracked in the PAC formulation, tracking path attributes would be challenging. Finally, the PAC approach has been applied to evaluate the vulnerability of system flow in the case where worst-case disruptions (maximization of flow loss) are concerned. Development of a similar model structure for identifying the set of facilities involved in the best-case disruption (minimization of system flow loss) is an important direction for future research.

**Acknowledgement**

This material is based upon work supported by the National Science Foundation under Grant No. 0720989. Any opinions, findings, and conclusions or recommendations expressed in this material are those of the author(s) and do not necessarily reflect the views of the National Science Foundation.

**References**

1. U.S. Congress, Office of Technology Assessment. Physical vulnerability of electric system to natural disasters and sabotage. OTA-E-453, Washington, DC: U.S. Government Printing Office; 1990.

2. Parfomak PW. Pipeline security: An overview of federal activities and current policy issues. Congressional Research Service Report for Congress; 2004.

3. Ambs K, Cwilich S, Deng M, Houck DJ, Lynch DF, Yan D. Optimizing restoration capacity in the AT&T network. Interfaces 2000; 20:26-44.

4. Cunningham WH. Optimal attack and reinforcement of a network. Journal of the Association for Computing Machinery 1985; 32(3):549-561.

5. Sohn J, Kim TJ, Hewings GJD, Lee JS, Jang S-G. Retrofit priority of transport network links under an earthquake. Journal of Urban Planning & Development 2003; 4:195-210.

6. Salmeron J, Wood K, Baldick R. Analysis of electric grid security under terrorist threat. IEEE Transactions on Power Systems 2004; 19(2):905-912.

7. Ham H, Kim TJ, Boyce D. Assessment of economic impacts from unexpected events with an interregional commodity flow and multimodal transportation network model. Transportation Research A 2005; 39(10):849-860.

8. Scaparra MP, Church RL. A bilevel mixed-integer program for critical infrastructure planning. Forthcoming in Computers & Operation Research 2006.

9. Church RL, Scaparra MP. Protecting critical assets the r-interdiction median problem with fortification. Geographical Analysis 2007; 39(2):129-146.

10. Bryson K-M, Millar H, Joseph A, Mobolurin A. Using formal MS/OR modeling to support disaster recovery planning. European Journal of Operational Research 2002; 141:679-688.

11. Kim TJ, Ham H, Boyce DE. Economic impacts of transportation network changes: Implementation of a combined transportation network and input-output model. Papers in Regional Science 2000; 81:223-246.

12. Fulkerson DR, Harding GC. Maximizing the minimum source-sink path subject to a budget constraint. Mathematical Programming 1977; 13:116-118.

13. Golden B. A problem in network interdiction. Naval Research Logistics Quarterly 1978; 25:711-713.

14. Corley HW, Sha DY. Most vital links and nodes in weighted networks. Operations Research Letters 1982; 1(4):157-160.

15. Ball MO, Golden BL, Vohra RV. Finding the most vital arcs in a network. Operations Research Letters 1989; 8(2): 73-76.

16. Israeli E, Wood RK. Shortest-path network interdiction. Networks 2002; 40(2): 97-111.

17. Church RL, Scaparra MP, Middleton RS. Identifying critical infrastructure: The median and covering facility interdiction problems. Annals of the Association of American Geographers 2004; 94(3):491-502.

18. Wollmer R. Removing arcs from a network. Operations Research 1964; 12:934-940.

19. McMasters AW, Mustin TM. Optimal interdiction of a supply network. Naval Research Logistics Quarterly 1970; 17:261-268.

20. Corley HW, Chang H. Finding the n most vital nodes in a flow network. Management Science 1974; 21:362-364.

21. Ratliff DH, Sicilia GT, Lubore SH. Finding the n most vital links in flow networks. Management Science 1975; 21(5):531-539.

22. Wood KR. Deterministic network interdiction. Mathematical and Computer Modelling 1993; 17(2):1-18.

23. Cormican KJ, Morton DP, Wood RK. Stochastic network interdiction. Operations Research 1998; 46(2):184-197.

24. Myung Y-S, Kim H. A cutting plane algorithm for computing k-edge survivability of a network. European Journal of Operational Research 2004; 156:579-589.

25. Murray AT, Matisziw TC, Grubesic TH. Critical network infrastructure analysis: Interdiction and system flow. Journal of Geographical Systems 2007; 9(2):103-117.

26. Matisziw TC, Murray AT, Grubesic TH. Bounding network interdiction vulnerability through cutset identification. Forthcoming In: Murray AT, Grubesic TH (Eds). Reliability and vulnerability in critical infrastructure: A Quantitative Geographic Perspective. Springer-Verlag, Advances in Spatial Science. 2007.

27. Ramos ER, Exposito AG, Santos JR, Iborra FL. Path-based distribution network modeling: Application to reconfiguration for loss reduction. IEEE Transactions on Power Systems 2005; 20(2):556-564.

28. Densham PJ, Rushton G. A more efficient heuristic for solving large p-median problems. Papers in Regional Science 1996; 71:307-329.

29. Matisziw TC. Modeling transnational surface freight flow and border crossing improvement. Dissertation, The Ohio State University, USA, 2005.

30. Matisziw TC, Murray AT, Grubesic TH. Evaluating vulnerability and risk in interstate highway operation. Proceedings of the Transportation Research Board Annual Meeting. Washington, DC. 2007.

31. Sorenson PA, Church RL. A comparison of strategies for data storage reduction in location-allocation problems. Geographical Systems 1996; 3:221-242.

32. Church RL. BEAMR: An exact and approximate model for the p-median problem. Computers & Operations Research 2008; 35(2):417-426.